\documentclass{Interspeech2024}
\usepackage{comment}
\usepackage{multirow, adjustbox}
\usepackage{amsmath,bm}
\usepackage{amssymb}
\usepackage{algorithm}
\usepackage{algpseudocode}
\usepackage{pifont}
\usepackage{xcolor}
\usepackage{hyperref}

\interspeechcameraready

\ninept
\title{All Neural Low-latency Directional Speech Extraction}

\name{Ashutosh}{Pandey}
\name{Sanha}{Lee}
\name{Juan}{Azcarreta}
\name{Daniel}{Wong}
\name{Buye}{Xu}

\address{
  Meta Platforms Inc., USA
  }
\email{\{apandey620, slee22, jazcarretao, ddewong, xub\}@meta.com}
\usepackage{textcomp}


\begin{document}

\maketitle

\begin{abstract}
    
    We introduce a novel all neural model for low-latency directional speech extraction. The model uses direction of arrival (DOA) embeddings from a predefined spatial grid, which are transformed and fused into a recurrent neural network based speech extraction model. This process enables the model to effectively extract speech from a specified DOA. Unlike previous methods that relied on hand-crafted directional features, the proposed model trains DOA embeddings from scratch using speech enhancement loss, making it suitable for low-latency scenarios. Additionally, it operates at a high frame rate, taking in DOA with each input frame, which brings in the capability of quickly adapting to changing scene in highly dynamic real-world scenarios. We provide extensive evaluation to demonstrate the model's efficacy in directional speech extraction, robustness to DOA mismatch, and its capability to quickly adapt to abrupt changes in DOA. 
    
    % This paper introduces a directional speech enhancement model that operates with an ultra-low latency of 2ms and processes raw audio waveforms directly. The model is adept at handling multi-talker conversation scenarios, where participants take turns speaking. The system is built on a spatial filter followed by a causal LSTM, with its hidden state conditioned on directional embeddings derived from prior knowledge of the desired speaker's spatial location. The model is trained using a target-switching mechanism, enabling it to adapt swiftly to sudden changes in spatial information. The system's effectiveness is demonstrated in a low-compute (600k parameters), low-latency (2ms) multi-talker scenario with an SNR of -5dB. The model's performance is compared with oracle beamformers and speech enhancement baselines that also utilize directional information, showcasing its superior capabilities. Furthermore, the model's robustness to errors in directional information is also demonstrated, underscoring its practical applicability.
\end{abstract}

\section{Introduction}

%In our increasingly digital world, clear and effective communication is paramount. However, speech signals, whether they are used in telecommunication, voice assistants, or other audio applications, are often degraded by interfering signals. This degradation can significantly impact the intelligibility and quality of the speech, leading to poor user experience and reduced system performance.

%Speech enhancement aims at improving the intelligibility and quality of target speech by attenuating interfering signals. Interference may consist of background noise, room reverberation, overlapping speech from interfering talkers, and echo \cite{loizou2013speech}. Speech enhancement proves beneficial for both human listeners and machines. Inspired by how humans process sound with two ears, multichannel speech enhancement uses audio signals from multiple microphones. This method allows it to gather spatial information about the target sound and any interfering sources in addition to the spectral-temporal information already present in the mixed sound \cite{gannot2017consolidated}.

In our digital-centric world, the clarity of speech signals is crucial. Yet, these signals, used in various applications, often suffer from interference, impacting their quality and user experience.
Speech enhancement, aiming to improve speech quality by reducing interference like background noise and reverberation \cite{loizou2013speech}, is beneficial for both humans and machines, such as automatic speech recognition. Multichannel speech enhancement uses multiple microphones to collect spatial and spectral-temporal information about the target and interfering sounds, enhancing the overall sound quality \cite{gannot2017consolidated}.

In real-world situations, distinguishing the target speaker amidst interference can be quite complex, often necessitating a two-step approach: speaker selection and speaker extraction. For humans, this selection is guided by attention and intention. In machine separation, current methods either separate all speakers from a mixture or use a cueing signal for speaker selection, followed by speaker extraction. These processes are respectively known as speaker separation and target speaker extraction.

Speaker separation, which isolates speakers from mixed audio, contends with the permutation ambiguity problem, where speaker-to-output stream allocation is ambiguous. Two key strategies, permutation invariant training (PIT) \cite{kolbaek2017multitalker} and deep clustering \cite{hershey2016deep}, address this issue. PIT's simplicity has notably driven significant progress in speaker separation \cite{luo2019conv, luo2020dual, chen2020dual, wang2023tf, subakan2021attention, quan2024spatialnet}.

Target speech extraction isolates a single speaker from a mixture using an auxiliary cue, ideal for single-output stream applications \cite{zmolikova2023neural}. These cues can be audio \cite{delcroix2018single, wang2018voicefilter, xu2020spex, li2020atss, zhang2020x, wang2021neural}, visual \cite{ephrat2018looking, afouras2018conversation, li2020listen}, spatial \cite{gu2019neural, brendel2020unified, kovalyov2023dsenet}, speech activity \cite{delcroix2021speaker}, or onset \cite{hao2021wase, pandey2023attentive}. 

With multichannel audio, the direction of arrival (DOA) is a key cue for identifying the target talker. This task, known as directional speech extraction (DSE), isolates speech from a fixed \cite{tesch2022insights,kovalyov2023dsenet, yang2024binaural} or adjustable \cite{chen2018multi, gu2019neural, brendel2020unified, xu2020neural, gu2020temporal, tesch2023multi} spatial region. As smart glasses and augmented/virtual reality technologies continue to advance, the importance of DSE is becoming increasingly evident. In addition to multichannel audio, these devices offer valuable signals like egocentric video and inertial measurement unit (IMU) data, assisting in determining the target talker's DOA.

\begin{figure*}[t]
  \centering
  \includegraphics[width=0.74\linewidth]{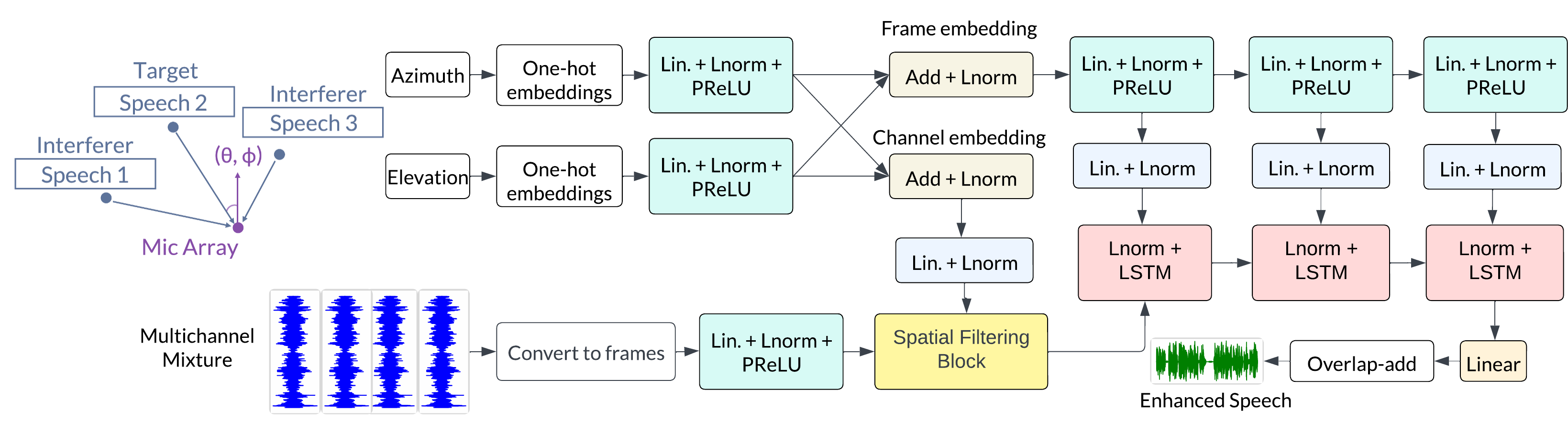}
  \caption{Schematic diagram of DRN.}
  \label{fig:drn_diag}
\end{figure*}

We introduce an all neural low-latency DSE model in this study. With end-to-end training, the model achieves a minimal algorithmic latency of $2$ ms. It uses DOA embeddings from a predefined spatial grid, transformed and integrated into a recurrent neural network (RNN) based speech extraction model, effectively isolating speech from the specified DOA.

Our method differs from previous works that used directional features like angular (DOA) feature \cite{chen2018multi}, directional power ratio, and directional signal to noise ratio to inform the model about DOA \cite{xu2020neural, gu2020temporal, zhang2021adl}. A similar approach was recently proposed in \cite{tesch2023spatially, tesch2023multi}, initializing RNN hidden states with DOA embeddings. However, this assumes a constant DOA for a given utterance. Our approach, using DOA for each input frame with a high frame rate, has better adaptability and is suited for highly dynamic real-world scenarios with moving sources and receivers.

Existing approaches to date primarily focus on Short-Time Fourier Transform (STFT) enhancement, which generally leads to higher latencies. In contrast, our model strategically employs a time-domain model to significantly reduce this latency. Although the study in \cite{kovalyov2023dsenet} proposed a low-latency filter-and-sum network (FaSNet) \cite{luo2019fasnet}, it is constrained by its ability to extract speech from a fixed spatial region. Our model surpasses this limitation, offering a more flexible solution.

We rigorously evaluate our model under a range of compute loads and latencies, and benchmark it against a strong baseline model and oracle beamformers. This comparison demonstrates improved speech enhancement performance metrics over these baselines in low-latency scenarios. Additionally, we assess the model's robustness to DOA estimation errors by introducing noise into the input DOA. Also, we illustrate the model's ability to rapidly adjust when the input DOA abruptly switches to a different speaker in the scene. This underlines the model's responsiveness to rapidly changing DOAs, a crucial feature for dynamic real-world scenarios. By improving multiple facets of DSE, our model represents a significant stride forward in this field.

\section{Proposed method}
\subsection{Problem formulation}

Consider a microphone array with $C$ channels and speech sources in a space with reverberation and background noise. In the time-domain, the observed multichannel mixture signal $\boldsymbol{Y} \in \mathbb{R}^{C \times N}$ can be decomposed as:

\begin{equation}
\boldsymbol{Y} = \boldsymbol{S}_d + \boldsymbol{S}_R + \boldsymbol{N}
\end{equation}

Here, $N$ is the number of time samples, $\boldsymbol{S}_d$ is the anechoic mixture of $K$ speakers, $\boldsymbol{S}_R$ includes room reverberation of speakers, and $\boldsymbol{N}$ contains additional interfering signals recorded by the array. Further, we can decompose the anechoic mixture as:

\begin{equation}
\boldsymbol{S}_d=\sum_{k=1}^K \boldsymbol{S}_{dk}
\end{equation}

where $\boldsymbol{S}_{dk}$ denotes the multichannel direct-path speech of talker $k$.

In this work, we aim to extract the speech signal $\boldsymbol{s}_{dkr}$ of a desired talker $k$ at a reference microphone $r$, using the noisy mixture $\boldsymbol{Y}$ and the DOA $\theta_k$ of the direct signal of talker $k$. Moreover, the desired talker might change at any timestamp, event that we refer to as \textit{target switching}.

\begin{figure*}[!t]
  \centering
  \includegraphics[width=0.85\linewidth]{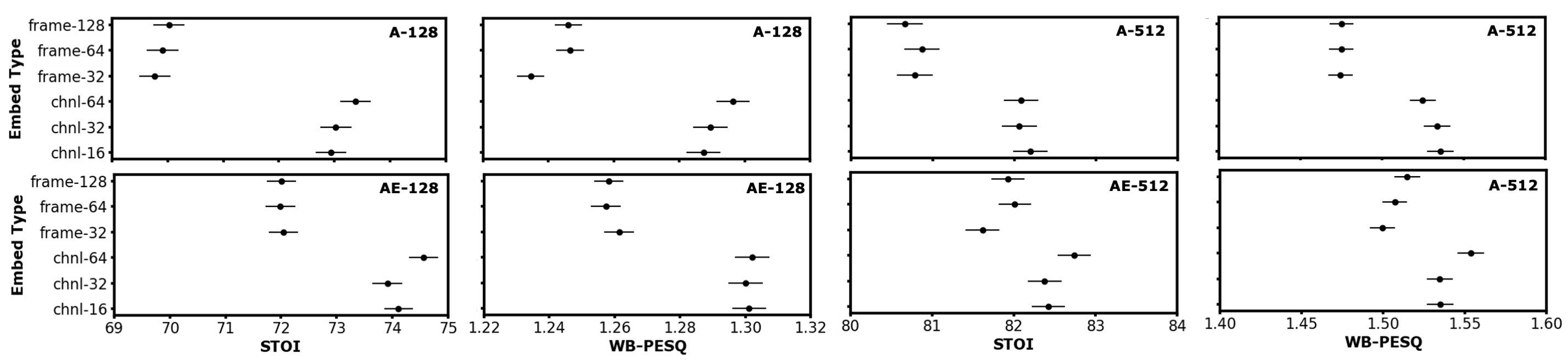}
  \caption{Comparing channel-wise and frame-wise embeddings. In plots labels, A and AE respectively represents azimuth-only and azimuth-elevation embeddings, whereas the number represents hidden size $H$.}
  \label{fig:objective_metrics_512_128}
\end{figure*}

\subsection{Model architecture}
We introduce Directional Recurrent Network (DRN), an all neural DOA-aware speech extraction network. DRN core design is based on the lightweight, low-compute and low-latency architecture introduced in \cite{pandey2023llrnn} for on-device time-domain speech enhancement. In this model, input waveform is initially converted into overlapping frames using a small frame shift of $R$ and an input window size of $iW$. The input is padded with $iW$ - $R$ zeros to construct the first frame. Subsequently, all frames undergo processing through trainable filters for spatial processing, followed by a stack of causal LSTM layers for temporal processing. At the output, enhanced frames of size $oW$ are estimated, and Overlap-Add (OLA) is applied to produce an enhanced waveform with an algorithmic latency of $oW$. %The model uses a small hop size (high frame rate) with a small output window to achieve low-latency \cite{wang2022stft}. 
 %The network output is projected into an output frame of size $oW$ samples, from where the time-domain signal is reconstructed by applying the Overlap-Add (OLA) method \cite{DSP3rded}. 
% Further, its compute can be changed quadratically with a single parameter $H$, the hidden size of the LSTM. For more details, refer to the original paper \cite{pandey2023llrnn}.

To condition this network on directional information, we define two types of DOA embeddings: channel-wise and frame-wise embeddings, which are extracted from DOA and injected at different stages of the network, as illustrated in Figure \ref{fig:drn_diag}.

The DoA information is grouped into $D$ directional bins and represented as one-hot vectors. The bin size defines the spatial grid resolution, which can be tuned to control the system memory depending on the application. The proposed method processes azimuth $\phi$ (horizontal) and elevation $\theta$ (vertical) information separately, and in the general case $D_{\phi} \neq D_{\theta}$.

\subsubsection{Channel-wise DOA embeddings}
We project time-dependent (frame-dependent) one-hot encoded azimuth and elevation vectors of size $T \times D_{\phi}$ and $T \times D_{\theta}$ independently using $C$ separate linear layers with layer norm and PReLU to obtain outputs of size $C \times T \times E_{C}$. Here, $T$ is the number of frames. Finally, azimuth and elevation streams are added together followed by layer norm to get the final channel-wise embeddings. Channel-wise embedddings are fused within the spatial processing blocks.

%The multichannel audio frame of shape $C \times T \time iW$ is projected to size $H$ using an inpuyt l;We fuse channel-wise DOA information with a signal of size multichannel audio using element-wise multiplication. After this, it is aggregated using the spatial processing block used in \cite{pandey2023llrnn}, which resembles frequency-dependent filtering.

\subsubsection{Frame-wise DOA embeddings}
For frame-wise embedding, we use a similar approach, but instead of using $C$ separate linear layers, we use one linear layer to obtain azimuth and elevation embedding of size $T \times E_{f}$. Frame-wise embeddings are fused after each of the LSTMs. 

\subsubsection{Fusing DOA embeddings}
Multichannel frames of size $C \times T \times iW$ are projected to size $H$ using a linear layer followed by layer norm and PReLU to obtain the input for spatial processing block of shape $C \times T \times H$.

The final channel-wise embeddings are projected to size $H$ using a linear layer and layer norm and then multiplied to the input of the spatial processing block. The final frame-wise embeddings go through a fully connected network with three hidden layers, each comprising a linear layer followed by layer norm and PReLU. The output from each of the hidden layers is projected using linear layer and layer norm to size $H$ and then multiplied to the corresponding LSTM output (see Figure \ref{fig:drn_diag}).

%The frame-wise location one-hot encoding vectors of shape $D \times T$ are projected into an embedding space of shape $E_f \times T$ to condition temporal location information to the temporal processing unit of DRN. This step is crucial to enable the model to quickly adapt when the location of the target talker changes after a \textit{target switching} event. As usual, we apply layer normalization and PreLu activation along the embedding dimension to the frame-wise embedding. Note that we create different location embeddings for each RNN layer.

%The frame-wise embedding is projected through a linear layer to match the input size $H$ of the RNN layers in DRN, and the frame-wise embeddings are fused to the input of each RNN layer by applying element-wise multiplication, followed by Layer Normalization and PreLu activation.

\begin{figure*}[tbp]
  \centering
  \includegraphics[width=0.92\linewidth]{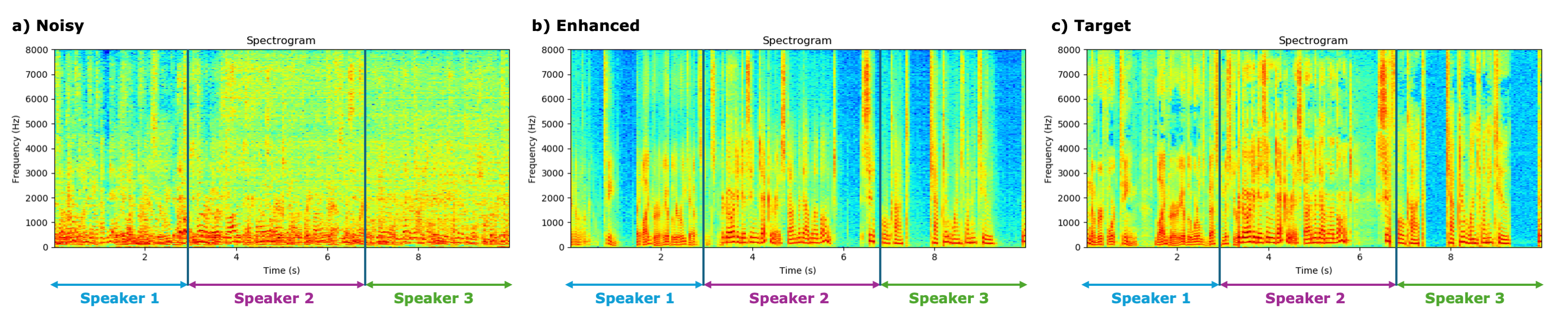}
  \caption{The spectrograms of a) the noisy audio, b) the DRN enhanced audio and c) the target audio. The input DOA of different talkers are switched to extract corresponding switched talkers at output.}
  \label{fig:figure_spectrum}
\end{figure*}

\begin{table}[!b]
    \centering
    \begin{adjustbox}{width=0.8\columnwidth}
    \begin{tabular}{|c|c|c|c|c|c|c|c|} 
    \hline
         $H$ &Lat. &Emb. &Domain&  STOI&  WB-PESQ&  SI-SDR& SNR\\ \hline 
         \multirow{6}{*}{ 128 } &\multirow{2}{*}{ 2ms }&A&T&  74.3&  1.31&  1.2& 4.2\\ 
         %128&a 2.5deg & 2 ms&T&  74.109&  1.301&  1.103& 4.145\\ 
          &&AE&T&  75.1&  1.31&  1.4& 4.3\\  
         \cline{2-8}
         %128&ae 2.5deg & 2 ms&T&  75.314&  1.316&  1.623& 4.366\\
  &\multirow{4}{*}{ 16 ms }& A & T& 74.6& 1.34& 1.7&4.6\\
 %128& a 2.5deg & 16 ms& T& 75.345& 1.351& 1.716&4.494\\
  && AE & T& 75.8& 1.36& 2.0&4.6\\
  && A & F& 74.7& 1.32& 1.3&4.3\\
 %128& a 2.5deg & 16 ms& F& 75.282& 1.338& 1.457&4.324\\
  && AE & F& 76.4& 1.36& 1.9&4.5\\
 %128& ae 2.5deg & 16 ms& F& 76.503& 1.370& 1.970&4.576\\ 
 \hline
 \hline
         \multirow{6}{*}{ 512 } &\multirow{2}{*}{ 2ms }&A &T&  82.3&  1.54&  4.6& 6.2\\ 
         %512&a 2.5deg & 2 ms&T&  82.699&  1.552&  4.746& 6.340\\ 
          &&AE &T&  83.1&  1.57&  4.9& 6.4\\ 
         \cline{2-8}
         %512&ae 2.5deg & 2 ms&T&  83.254&  1.573&  4.985& 6.473\\
  &\multirow{4}{*}{ 16 ms }& A & T& 84.8& 1.69& 6.0&7.2\\
 %512& a 2.5deg & 16 ms& T& 85.003& 1.687& 5.974&7.205\\
  && AE & T& 85.5& 1.71& 6.2&7.3\\
 %512& ae 2.5deg & 16 ms& T& 85.991& 1.729& 6.436&7.516\\
  && A & F& 84.3& 1.63& 5.6&6.9\\
 %512& a 2.5deg & 16 ms& F& 84.665& 1.675& 5.809&7.078\\
  && AE & F& 85.5& 1.70& 6.1&7.3\\
 %512& ae 2.5deg & 16 ms& F& 85.785& 1.725& 6.230&7.373\\ 
 \hline
    \end{tabular}
\end{adjustbox}
    \caption{Comparing time and frequency domain DRN model for two embedding types.}
    \label{results_table}
\end{table}
\begin{table}[!t]
\centering
\begin{adjustbox}{width=\columnwidth}
\begin{tabular}{|c|c|c|c|c|c|c|}
\hline
Model & Domain & Lat & Hidden size & STOI & WB-PESQ & SI-SDR \\
\hline
\multicolumn{4}{|c|}{Noisy} & 49.2 & 1.09 & -13.08 \\
\hline
\multirow{4}{*}{ DRN } & \multirow{2}{*}{ FD } &  \multirow{2}{*}{ 16 ms } & 128 & 76.5 & 1.37 &  1.96 \\
& & & 512 & \textbf{85.7} & \textbf{1.72} & \textbf{6.2}   \\
\cline{2-7}
& \multirow{2}{*}{ TD } &  \multirow{2}{*}{2 ms}  & 128 & 75.3 & 1.31 & 1.62  \\
& & & 512 & 83.2 & 1.57 & 4.98  \\
\hline
\multirow{2}{*}{ Oracle MCWF } & \multirow{2}{*}{ FD } & 16 ms & - & 77 & 1.26 & 3.4 \\
& & 2 ms & - & 62.5 & 1.16 & 1.9  \\
\hline
\multirow{4}{*}{ maxDI informed }& \multirow{2}{*}{ FD } &  \multirow{2}{*}{ 16 ms } & 128 & 77.3 & 1.37 & 1.78  \\
& & & 512 & 84.7 & 1.67 &  5.29 \\
\cline{2-7}
& \multirow{2}{*}{ TD } &  \multirow{2}{*}{2 ms}  & 128 & 70.3 & 1.25 & -0.82 \\
& & & 512 & 80.9  & 1.48 & 3.64 \\
\hline
\end{tabular}
\end{adjustbox}
\caption{Comparing DRN with spatial filtering baselines for frequency-domain (FD) and time-domain (TD) processing.}
\label{beam_results_table}
\end{table}

\section{Dataset Generation}
We use the Interspeech2020 DNS Challenge corpus \cite{reddy2020icassp} to generate pairs of clean and noisy signals as in \cite{pandey2023llrnn}. An $8$-microphone circular array with a radius of $10$ cm is used to create multichannel data. RIRs are generated for shoebox like room geometries by using the Image Source Method of order $6$. We generate $10$ seconds long $160$k training, $500$ validation, and $3.2$K test utterances with a sampling rate of 16kHz.

We first sample room dimension of size $[L, W, H]$ in meters, where $L$ is sampled from $[3, 10]$, $W$ from $[3, 10]$ and $H$ from $[2, 5]$. Further simulation conditions are sampled as follows: wideband absorption coefficients from $[0.1, 0.4]$, random array location and orientation at least 0.3m away from the room walls, and $1$ to $5$ target talker locations at a distance $[0.5, 2.5]$ meters from array. The azimuth separation between target talkers is kept to be at least $20$ degrees. With a probability of $0.75$, $1$ to $10$ interfering talkers are added at least $3$ meters from array to simulate cases of distant speech interference and babble noise. Additionally, $1$ to $10$ noise (non-speech) sources are added at least 0.5 meters from the array. All the sources are set to be at least $0.3$ meters from wall. 

The RMS level of target talkers and noises are sampled from $[-2.5, 2.5]$ dB. The RMS level of interfering talkers are sampled from $[-10, -5]$ dB. Interfering talkers are added with an SIR from $[5, 10]$ dB  and noises are added  with an SNR from $[-5, 10]$ dB. Signal ratios are calculated with respect to the direct path signal of the quietest target talker in the clip. When scaling interfering talkers, a scale greater than $1$ is set to $1$ to avoid target-interference confusion. 

The target signal may contain non overlapping regions of multiple talkers. This is achieved by applying \textit{target switching}. The input DOA stream contains DOA of corresponding talker in the target. We apply either no switching, one switch or two switch within a 10 seconds long utterance, without exceeding the number of target talkers in the clip. During training, the switching time is randomized, while during testing switching occurs at uniform intervals of $\textit{clip\_length} /\{\# of switches + 1\}$. During training, the switching interval uniformly deviates $[-5, +5] \%$ of utterance length from the uniform interval points. The training target is set to be the direct-path speech
at the first microphone (r = 1).

\section{Results}

\subsection{Experiments design}
We developed all models in Pytorch using automatic mixed precision training and phase constrained magnitude (PCM) loss, similar to the model in \cite{pandey2023llrnn}. Training parameters included a batch size of 16 utterances, Adam optimizer with amsgrad, a learning rate of 0.0002, and gradient norm clipping to 0.03, with truncated backpropagation over 100 epochs. During training, temporal jitter was introduced to the input DOA stream by adding uniform noise, ranging from [-2.5, 2.5] degrees, to a sampled mean DOA deviation from the ground truth from [-2.5, 2.5]. Model comparisons were made using short-time objective intelligibility (STOI), wide-band perceptual evaluation of speech quality (WB-PESQ), SNR, and SI-SDR.

\subsection{Finding the optimal location embeddings}
To determine the optimal location embedding size and type, we compared the performance of a DRN model with $2$ ms latency using azimuth-only (A), and azimuth-elevation (AE) DOA embeddings. We tested three channel-wise embeddings sizes ($16$, $32$, $64$) and three frame-wise embeddings sizes ($32$, $64$, $128$). We also compared these settings' performance for two different hidden sizes ($H=128$ and $H=512$). These results with their corresponding confidence intervals are plotted in Figure \ref{fig:objective_metrics_512_128}.

We note that in all instances, channel-wise embeddings outperform frame-wise embeddings. As anticipated, AE performs better than A, as AE aids in vertical source localization as well. Interestingly, an increase in embedding sizes does not consistently result in performance improvement.

Next, we train the model with combined channel-wise and frame-wise embeddings. We used the best embeddings size determined from Figure \ref{fig:objective_metrics_512_128}. The objective scores are shown in Table \ref{results_table}. For both $H=128$ and $H=512$ cases, AE is better.  

%Once we identified the best location embedding size, we trained the model with the combined channel and the frame embeddings. We used the best embeddings size determined from the previous experiment. For this combination experiment we tested the azimuth only, azimuth and elevation cases. We again trained models using both the hidden size $H=128$ and $H=512$. The objective metrics scores calculated are shown in Table \ref{results_table}.  

\begin{figure}[!t]
  \centering
  \includegraphics[width=0.6\linewidth]{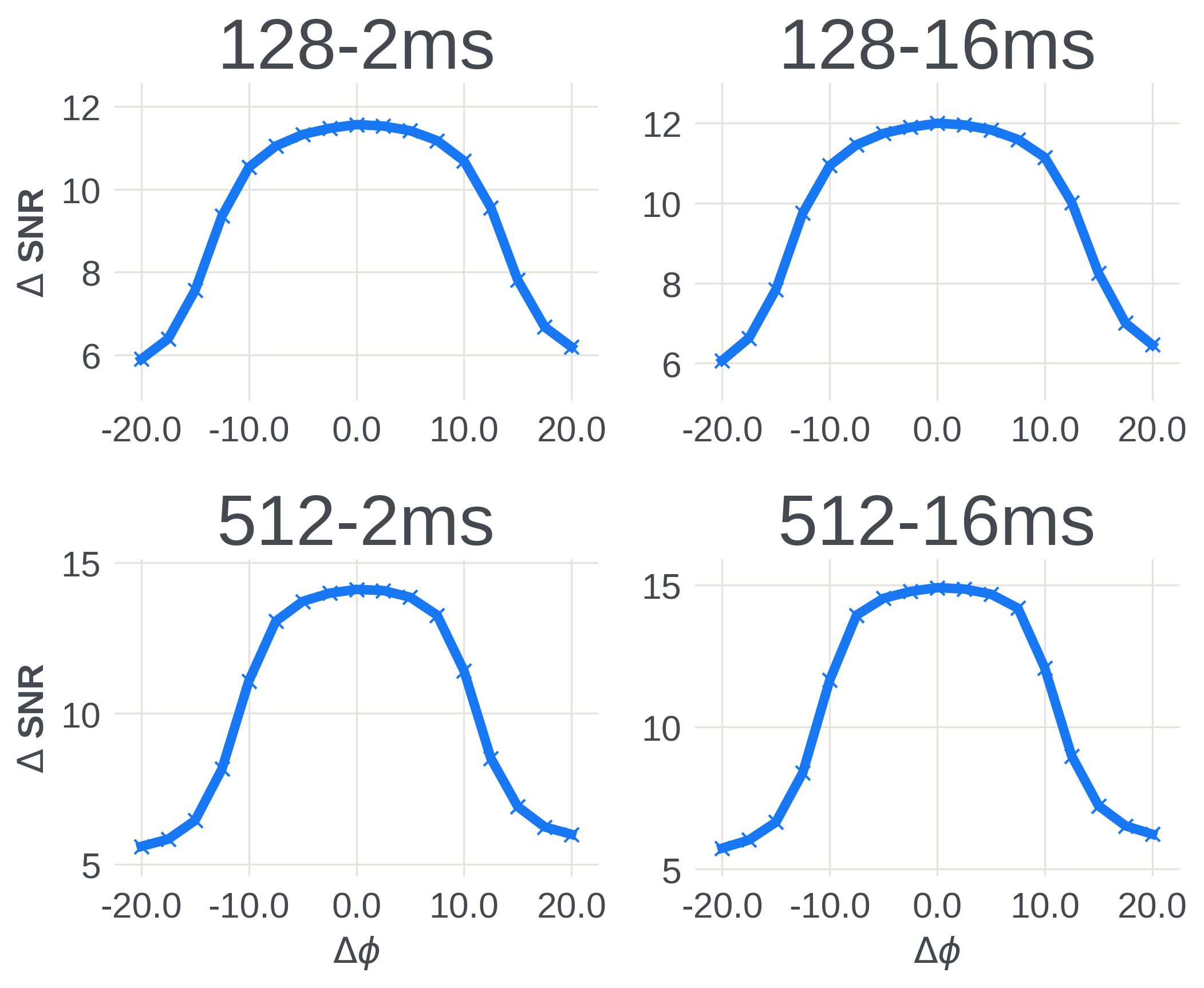}
  \caption{Sensitivity of DRN to DOA mismatch.}
  \label{fig:drn_sensitivity_analysis}
\end{figure}

\begin{table}[!b]
\centering
\begin{adjustbox}{width=1\columnwidth}
\begin{tabular}{|c|c|c|c|c|c|c|}
\hline
Model & Hidden Size & Latency & STOI & WB-PESQ & GMAC/s & \#Params \\
\hline
\multicolumn{3}{|c|}{Noisy} & 51.3 & 1.11 & - & - \\
\hline
\multirow{6}{*}{ DRN } & \multirow{2}{*}{ 256 } & 2 ms & 80.3 & 1.47 & 2.3 & 1.8 \\
& & 32 ms & 81.8 & 1.59 & 2.9 & 2.0 \\
\cline{2-7}
& \multirow{2}{*}{ 512 } & 2 ms & 83.2 & 1.59 & 7.8 & 6.7 \\
& & 32 ms & 86.4 & 1.79 & 9.0 & 7.1 \\
\cline{2-7}
& \multirow{2}{*}{ 1024 } & 2 ms & 86.2 & 1.72 & 28.3 & 26.0 \\
& & 32 ms & \textbf{89.3} & \textbf{2.00} & 30.5 & 26.8 \\
\hline
McNet \cite{tesch2023multi} & - & 32 ms & 85.8 & 1.72 & 30.2 & 1.9 \\
\hline
\end{tabular}
\end{adjustbox}
\caption{Comparing DRN with a DNN based strong DSE model.}
\label{mcnet_results_table}
\end{table}

\subsection{Higher latency and frequency domain processing}

We trained  DRN with hidden sizes $H=128$ and $H=512$, embedding classes A and AE, and 16 ms latency in both time and frequency domains. For the frequency domain, time-domain frames were replaced with the corresponding real and imaginary components of DFT. In all cases, AE performed better than A. Performance significantly improved for both hidden sizes when latency increased from $2$ ms to $16$ ms. However, transitioning from a $16$ ms time domain to a 16 ms frequency domain did not yield a notable performance improvement.

\subsection{Switching the direction of the target speech}

A key question in this study is whether the model can swiftly adapt when the input DOA switches to a different talker. Figure \ref{fig:figure_spectrum} shows that when speakers switch, the spectrogram of the enhanced signal aligns well with the target signal, suggesting successful target switching and expected speech extraction by the model. Listening tests further validate the model's ability to rapidly switch between talkers.

\subsection{Baseline models}
In Table \ref{beam_results_table} we compare the best DRN model of each category with spatial signal processing baselines. For these experiments the azimuth and elevation grid resolution is 2.5 degrees.
First, we show that DRN can outperform oracle Multichannel Wiener Filter (MCWF) \cite{gannot2017consolidated}, where both the multichannel desired speech and the target-switching timestamps are available. We perform causal MCWF in the Frequency-Domain and update the covariance matrices in an online fashion by running a cumulative sum over time that resets its state at a target switching event. Latency of MCWF is controlled with the window size and by setting the hop size as half of the window value.

We also created a maxDI informed baseline by feeding to the model in \cite{pandey2023llrnn} the microphone audio along with their corresponding signals filtered by a DOA-aware maximum Directivity Index (maxDI) beamformer. %The maxDI beamformer suppresses diffuse noise without introducing distortions to the speech arriving from the input DOA.%
The front-end maxDI runs in the frequency-domain, and to keep the system latency unchanged the beamformer delay is not compensated for the time-domain models. It is interesting to note that in Table \ref{beam_results_table}, the FD DRN model with size $H=128$ does not show a clear improvement over the maxDI informed baseline. However, for all other cases, the DRN model outperforms the maxDI informed baseline.

Finally, we compared DRN with a strong frequency-domain DNN-based model, MCNet, proposed in \cite{tesch2023multi}. As MCNet cannot accept time-dependent DOA, we trained both DRN and MCNet without DOA switching. The results, presented in Table \ref{mcnet_results_table}, show that DRN achieves superior results with significantly lower computation for $H=512$. Interestingly, to surpass MCNet with a DRN model at $2$ ms latency, we had to increase $H$ to $1024$ to match MCNet's computation.

\subsection{Robustness to DOA Mismatch}
Lastly, we assessed DRN's performance when DOA gradually deviates from the ground truth. In this case, DRN was trained with azimuth-only embeddings at a resolution of $2.5$ degrees. We calculated the average improved SNR ($\Delta$ SNR) over $3200$ single-talker mixtures with $1-10$ noise sources, when DOA shifted from the ground truth from $[-20.0, -17.5, \cdots, 0, \cdots, 17.5, 20.0]$. Results are plotted in Figure \ref{fig:drn_sensitivity_analysis}. DRN showed a flat response close to the true DOA up to + and - $5$ degrees and a drop within 0.7 dB within + and - $10$ degrees of the ground truth, demonstrating its robustness to DOA estimation errors.

\section{Conclusions}
We have presented a novel all neural model for directional speech extraction. The unique approach of using trainable DOA embeddings with an RNN based speech extraction model has offered a low-compute, low-latency solution. Through rigorous evaluations, we have demonstrated models effectiveness in speech extraction, its robustness to DOA mismatch and its ability to quickly adapt to abrupt changes in DOA. This research opens up new possibilities for further improvements and applications. Future work will focus on refining the model with training on data containing source and receiver movements.

\bibliographystyle{IEEEtran}
\bibliography{mybib}

\end{document}